\documentclass[12pt,fleqn]{iopart}
\usepackage{color}
\usepackage{amssymb}
\usepackage{amsmath}
\usepackage{mathtools}
\usepackage{graphicx}
\usepackage{epsfig}
\usepackage{stackengine}
\usepackage{braket}
\usepackage{makecell}
\usepackage{subfigure}
\usepackage{epsf}
\usepackage{epstopdf}
\usepackage{xfrac}
\usepackage[T1]{fontenc}
\def\delequal{\mathrel{\stackon[1pt]{=}{$\scriptstyle\Delta$}}}
\newtheorem{definition}{Definition}
\begin{document}

\title[Mixed States Robust Recovery]{Robust quantum state recovery from amplitude damping within a mixed states framework}

\author{Saeideh Shahrokh Esfahani, Zeyang Liao, and M. Suhail Zubairy}

\address{Institute for Quantum Science and Engineering (IQSE) and Department of Physics and Astronomy, Texas A$\&$M University, College Station, TX 77843-4242, USA}

\begin{abstract}
Due to the interaction with the environment, a quantum state is subjected to decoherence which becomes one of the biggest problems for practical quantum computation. Amplitude damping is one of the most important decoherence processes. 
Here, we show that general two-qubit mixed states undergoing an amplitude damping can be almost completely restored using a reversal procedure. This reversal procedure through CNOT and Hadamard gates, could also protect the entanglement of two-qubit mixed states, when it undergoes general amplitude damping. Moreover, in the presence of uncertainty in the underlying system, we propose a robust recovering method with optimal characteristics of the problem.
\end{abstract}


\section{Introduction}

In the process of inevitable interaction with the environment, the elements of a quantum computational system can entangle with the environment and consequently become decoherent. This decoherence procedure is a fundamental obstacle for successful transfer of quantum information and for practical quantum computation.
A number of effective approaches have been proposed to suppress the decoherence effect. One way for protecting a quantum state from decoherence is based on the existence of decoherence free subspaces of states which requires special symmetry properties of the interaction Hamiltonian. In Quantum Computation, this procedure, i.e. utilization of decoherence free subspaces of states, is called ``error-avoiding code'' \cite{lidar1998decoherence}. ``Quantum error correction code (QECC)'' is another way to suppress the decoherence effect~. In QECC, the logical quantum bit (qubit) is encoded in a larger Hilbert space of several physical qubits and the correction process is performed by constructing proper measurements and correction operations \cite{shor1995scheme, duan1997preserving}. Other methods include Quantom Zeno effect~\cite{facchi2004unification,maniscalco2008protecting} and dynamical decoupling~\cite{viola1998dynamical,viola1999dynamical} which have also been widely used to mitigate decoherence and to protect the quantum state.

Amplitude damping is an important type of decoherence which can happen in many quantum systems \cite{weisskopf1930calculation}, including a photon qubit in a leaky cavity, atomic qubit subjected to spontaneous decay, or a super-conduction qubit with zero-temperature energy relaxation. It can cause errors in quantum information transfer and quantum computation results. In the past years, several strategies have been proposed to protect quantum states from the amplitude damping. Three widely used strategies to protect the quantum state from amplitude damping are: (1) Weak measurement reversal~\cite{sun2010reversing,korotkov2006undoing,sun2009reversing}, (2) Quantum un-collapsing (reversal measurement) of the quantum state towards the ground state before the amplitude damping, which can largely suppress the decoherence~\cite{korotkov2010decoherence}, and (3) Utilization of quantum gates to restore a qubit state in a weak measurement ~\cite{al2011reversing, liao2013protecting}. Quantum state recovery based on quantum gates can be accomplished in a shorter time. It is shown that a one-qubit state in a weak measurement can be completely recovered by applying Hadamard and CNOT gates on the system qubit and an auxiliary qubit~\cite{al2011reversing}. This method is generalized to recover an arbitrary two-qubit pure state undergo amplitude damping \cite{liao2013protecting}. In this paper, we show that this method can also be used to protect an arbitrary two-qubit mixed states. Furthermore, we also consider the cases when there is uncertainty in the input density matrix, the damping parameter, or both. We test our recovery schemes by generating arbitrary mixed states density matrices via extensive Monte-Carlo simulations and show that the optimal solution can be fairly approximated by the original scheme when the parameters are replaced by their average values in the uncertainty space.

The paper is organized as follows: in Sec.~\ref{sec-damping}, we calculate the damped matrix of arbitrary two-qubit mixed state.  In Sec.~\ref{sec-recovery}, we demonstrate how we use the proposed recovery scheme to reverse the damping effect and recover the quantum state.  In section~\ref{sec-extended}, an extended scheme~\cite{liao2013protecting} is applied to amplify protection proposed in section~\ref{sec-concurrence}. In Sec.~\ref{sec-uncertainty} a robust recovery under uncertainty of the input states and damping parameters is studied. Finally we summarize the results.

\section{Amplitude damping of two-qubit mixed state}\label{sec-damping}

Amplitude damping is an important type of decoherence and a single qubit amplitude damping can be mathematically described by the following mappings:
\begin{eqnarray}
|0\rangle_{S}|0\rangle_{E} &\rightarrow & |0\rangle_{S}|0\rangle_{E} \nonumber \\
|1\rangle_{S}|0\rangle_{E} &\rightarrow & \sqrt{1-p}|1\rangle_{S}|0\rangle_{E} + \sqrt{p}|0\rangle_{S}|0\rangle_{E}
\end{eqnarray}
where $p\in [0,1]$ is the possibility of decaying of the excited state, and S (E) denotes the system (environment). Within the Weisskopf-Wigner approximation \cite{weisskopf1930calculation}, we have $\sqrt{1-p}=e^{-\Gamma t}$. For a general single qubit mixed state $\rho$, amplitude damping can also be written as
\begin{equation}
\rho\rightarrow \varepsilon _{AD}(\rho)=A_{0}\rho A_{0}+A_{1}\rho A_{1}
\end{equation}
where the amplitude damping operations are given by
\begin{equation}
{ A }_{ 0 }=\left(\begin{array}{cc}

 1 & 0 \\ 0 & \sqrt { 1-p }  \end{array} \right), { A }_{ 1 }=\left(\begin{array}{cc}

 0 & \sqrt { p } \\ 0 & 0   \end{array} \right). 
 \end{equation}
 
An arbitrary two-qubit mixed state can be written as 
\begin{equation}\label{eqn-initial}{ \rho  }_{ i }=\left( \begin{array}{cccc} a & e & f & g \\ { e }^{ * } & b & h & i \\ { f }^{ * } & { h }^{ * } & c & j \\ { g }^{ * } & { i }^{ * } & { j }^{ * } & d \end{array} \right).\end{equation}
The amplitude damping of an arbitrary two-qubit mixed state can be calculated by the following procedures. 
First an arbitrary two-qubit mixed state can be also written as $\rho=\sum_{i,j,m,n=0}^{1}\alpha_{ijmn}|ij\rangle\langle mn|$. Each element $|ij\rangle\langle mn\rangle$ can be written as two-qubit direct products $|i\rangle\langle m|\otimes |j\rangle\langle n|$. Next we apply the amplitude damping operations on each qubit which yields
\begin{equation}
|ij\rangle\langle mn| \rightarrow \left[ { A }_{ 0 }\ket {i} \bra{m} { A }_{ 0 }^{ \dagger  }+{ A }_{ 1 }\ket{i} \bra{m} { A }_{ 1 }^{ \dagger  } \right] \bigotimes \left[ { A }_{ 0 } \ket{j} \bra{n} { A }_{ 0 }^{ \dagger  }+{ A }_{ 1 } \ket{j} \bra{n} { A }_{ 1 }^{ \dagger  } \right].
\end{equation}
After applying the amplitude damping operations on each element, we obtain the two-qubit amplitude damped state given by
\begin{equation}\label{eqn-damped}{ \rho  }_{ d }=\left( \begin{array}{cccc}
 a+bp+cp+{ p }^{ 2 }d & e\sqrt { q } +pj\sqrt { q }  & f\sqrt { q } +ip\sqrt { q }  & gq \\ { e }^{ * }\sqrt { q } +{ j }^{ * }p\sqrt { q }  & bq+pdq & hq & iq\sqrt { q }  \\ { f }^{ * }\sqrt { q } +{ i }^{ * }p\sqrt { q }  & { h }^{ * }q & cq+pdq & jq\sqrt { q }  \\ { g }^{ * }q & { i }^{ * }q\sqrt { q }  & { j }^{ * }q\sqrt { q }  & d{ q }^{ 2 } \end{array} \right).\end{equation}
where $q=1-p$.


\section{Two-qubit mixed states recovery}\label{sec-recovery}

\begin{figure}
\begin{center}
\includegraphics[scale=0.6]{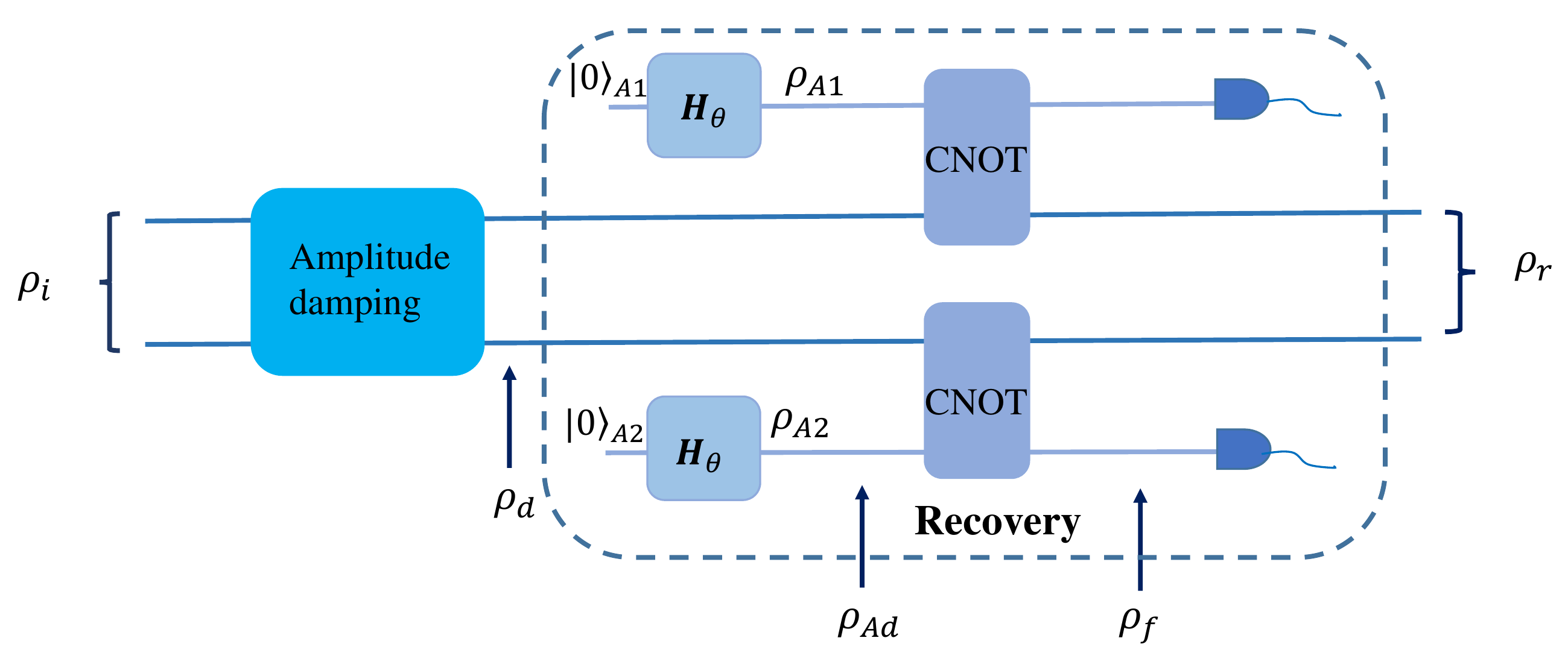}
\caption{A schematic view of the recovery process proposed in~\cite{liao2013protecting}, generalized herein for the mixed states setting.}
\label{fig-recovery}
\end{center}
\end{figure}

In this section, we propose a method to recover the damped quantum mixed states in Eq. \eqref{eqn-damped} to the initial quantum mixed states in Eq. \eqref{eqn-initial}. This method has recently been introduced for two-qubit pure state~\cite{liao2013protecting}. In this model, we use the circuit diagram outlined in Fig.~\ref{fig-recovery}. Two auxiliary qubits both in the $\ket{0}$ state initially, are added. First, we apply a Hadamard gate with angle $\theta$ for each ancilla qubit. 
\begin{equation}
 H_{\theta}=\left( \begin{array}{cc}
\cos\theta & -\sin\theta \\
\sin\theta & \cos\theta  \end{array} \right)
\end{equation}
The ancilla qubits (${ A1 }$ and ${ A2 }$) after passing through the Hadamard gate will change to:
\begin{equation}
{ { \rho  } }_{ A1 }={ { \rho  } }_{ A2 }=\left(\begin{array}{cc} { \cos { \theta  }  }^{ 2 } & \cos { \theta  } \sin { \theta  }  \\ \cos { \theta  } \sin { \theta  }  & { \sin { \theta  }  }^{ 2 } \end{array}\right).
\end{equation}
The state of the whole system, after combining ancilla qubits to the damped system in Eq. \eqref{eqn-damped} can be written as:
\begin{equation}
{ \rho  }_{ Ad }={ \rho  }_{ A1 }\bigotimes { \rho  }_{ d }\bigotimes { \rho  }_{ A2 }.
\end{equation}
Afterwards, we apply two CNOT gates onto each pair of the system and ancilla qubits:

 \begin{table}[h]
\centering
\begin{tabular}{c|c}
CNOT gate I &  CNOT gate II 
\\
\hline
${ UC }_{ 1 }=\left( \begin{array}{cccc} 1 & 0 & 0 & 0 \\
0 & 0 & 0 & 1\\
0 & 0 & 1 & 0 \\
0 & 1 & 0 & 0
\end{array}\right)$ & 
${ UC }_{ 2 }=\left( \begin{array}{cccc} 
1 & 0 & 0 & 0\\
0 & 1 & 0 & 0 \\
0 & 0 & 0 & 1\\
0 & 0 & 1 & 0
\end{array}\right)$
\end{tabular}
\caption{ CNOT gates used for recovery in Fig.~\ref{fig-recovery}}
\label{my-label}
\end{table}
The final state is then given by:
\begin{equation} { \rho  }_{ f }={ UC }_{ 1 }\bigotimes { UC }_{ 2 }\cdot
{ \rho  }_{ Ad }\cdot { { UC }_{ 2 } }^{ \dagger  }\bigotimes { { UC }_{ 1 } }^{ \dagger  }
\end{equation}
Finally, we make measurements on the two ancilla qubits. If the ancilla qubits are both in $\ket{0}$ state, the recover process is successful. Otherwise, the recover process fails. Since the ancilla qubits are measured to be in $\ket{0}$ state, the state of the whole system becomes
\begin{equation}
\rho_{f}^{'}=(P_{A1}\otimes P_{A2})\rho_{f}(P_{A2}^{\dagger }\otimes P_{A1}^{\dagger })
\end{equation}
where the projection operators $P_{ A1 }= \ket{0}\bra{0}\bigotimes \mathbb{I}_{2}$ and $P_{ A2 }=\mathbb{I}_{2} \bigotimes \ket{0}\bra{0}$ with $\mathbb{I}_{2}$ being a two-by-two unit matrix.
The reduced system density matrix is ${ \rho  }_{ r }={ Tr }_{ A1, A2 }({ \rho  }_{ f1 }^{'})$ where ${ Tr }_{ A1, A2 }$ denotes the partial trace over the ancilla qubits. 

By choosing $\theta=\tan^{-1}(1/\sqrt{q})$, the final state after the recovery process for the system can be calculated to be
  \begin{equation}\label{eqn-recovered}
{ \rho  }_{ r }=\quad \frac { 1 }{ 1+(1-a+d)p+{ p }^{ 2 }d } \left( \begin{matrix} a+bp+cp+{ p }^{ 2 }d & e+pj & f+ip & g \\ e^{*}+j^{*}p & b+pd & h & i \\ f^{*}+i^{*}p & h^{*} & c+pd & j \\ g^{*} & i^{*} & j^{*} & d \end{matrix} \right).   
  \end{equation}
Eq. ~\ref{eqn-recovered} can be rewritten as $\rho_{r}=(\rho_{i}+\rho_{err})/N$ where $\rho_{i}$ is the initial state, $N=1+(1-a+d)p+{ p }^{ 2 }d$ is the normalization factor, and $\rho_{err}$ the recovering error matrix which is given by
   \begin{equation}\label{eqn-recovered-simplify}
{ \rho  }_{ err }= \left( \begin{matrix} bp+cp+{ p }^{ 2 }d & jp & ip & 0 \\ j^{*}p & dp & 0 & 0 \\ i^{*}p & 0 & pd & 0 \\ 0 & 0 & 0 & 0 \end{matrix} \right).   
  \end{equation}
Thus, the system is not completely recovered but is restored to the initial input density matrix plus an error term. When $p=0$, $\rho_{r}=\rho_{i}$ which is expected. In the following subsections, we quantitatively analyze how the quantum state is restored using fidelity and quantum concurrence.

\subsection{Fidelity}\label{sec-fidelity}

One way to measure how a quantum state is recovered is by calculating the fidelity between the recovered and the initial states.  The fidelity function between two quantum mixed states is defined by~\cite{jozsa1994fidelity}
\begin{equation}\label{eqn-fidelity}
F(\rho_{i},\rho_{f})=\left[Tr\left(\sqrt{\sqrt{\rho_{i}}\rho_{f}\sqrt{\rho_{i}}}\right)\right]^{2},
\end{equation}
where $\rho_{i}$ and $\rho_{f}$ are the initial and final state, respectively.
In this paper, the fidelity between the damped state and the initial state is  ${ F }_{ d }= F({\rho}_{i},{\rho}_{d})$, and the fidelity between the recovered state and the initial state is  ${ F }_{ r }=F({\rho}_{i},{\rho}_{r})$. 

\begin{figure}
\begin{center}
\subfigure[Fidelity for density matrix $\rho_1$]{\label{fig-Fidelity1}
\includegraphics[scale=0.5]{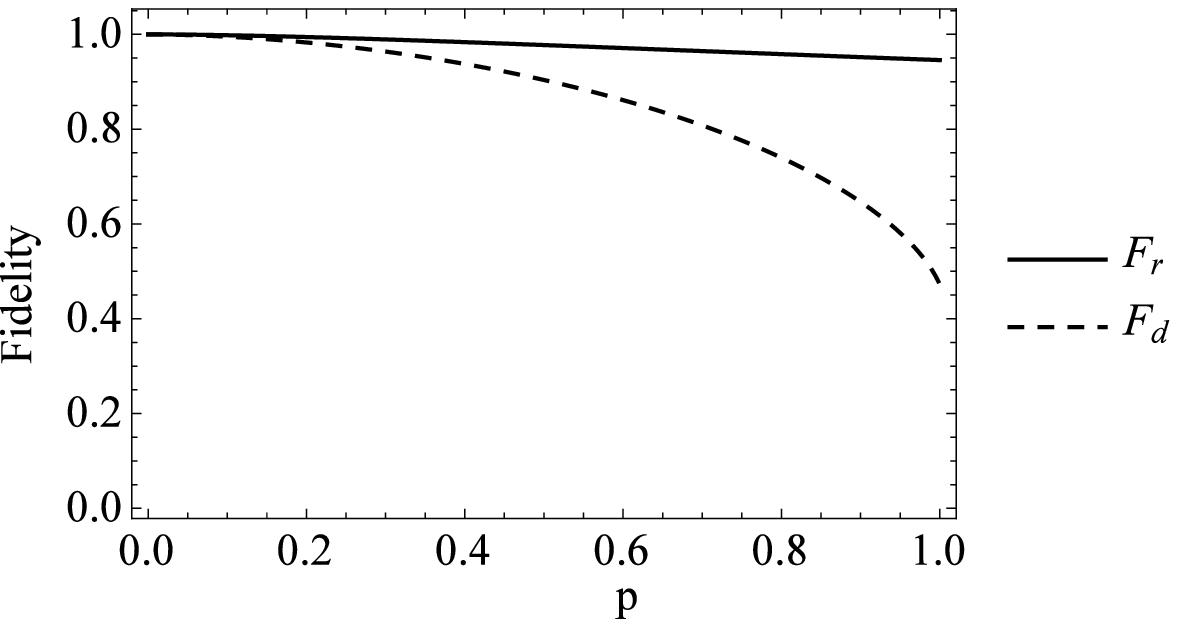}}
\subfigure[Fidelity for density matrix $\rho_2$]{\label{fig-Fidelity2}
\includegraphics[scale=0.5]{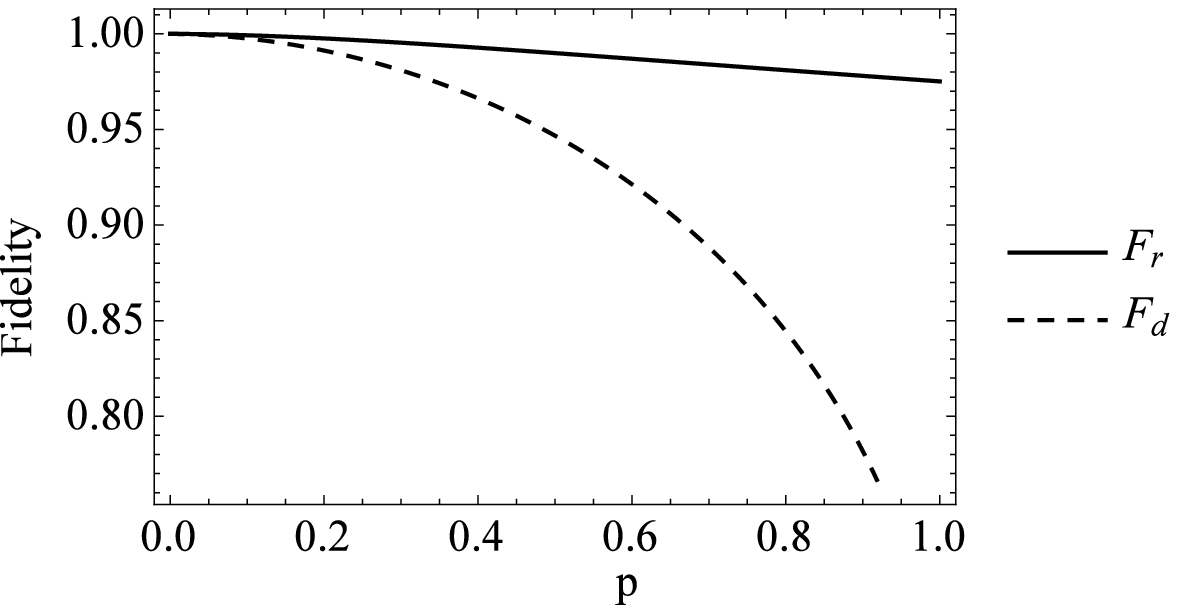}}
\caption{The fidelity between the initial and the final states. ${F}_{d}$ corresponds to the fidelity of recovered state and ${F}_{r}$ corresponds to recovered state. $\rho_1=\left( \begin{matrix} 0.4 & 0 & 0 &0.25 \\ 0 & 0.1 &0  & 0 \\ 0 & 0 & 0.3 & 0 \\ 0.25 & 0 & 0 & 0.2 \end{matrix} \right)$ , $\rho_2=\left( \begin{matrix} 0.6 & 0 & 0 & 0.25 \\ 0 & 0.12 & 0 & 0 \\ 0 & 0 &0.11& 0 \\ 0.25 & 0 & 0 &0.17 \end{matrix} \right)  $  }

\end{center}
\end{figure}

\begin{figure}
\begin{center}
\includegraphics[scale=0.5]{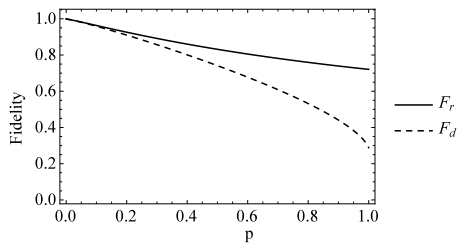}
\caption{The average fidelity of the damped and recovered states via Monte Carlo method with $10^{4}$ iteration as a function of damping probability ($p$)}
\label{fig-fidelity-Unknown}
\end{center}
\end{figure}

The recovering fidelities as a function of the decaying probability $p$ for two mixed states are illustrated in Fig.\ref{fig-Fidelity1} and \ref{fig-Fidelity2}. From the figures, we see that the fidelities of the recovered states are higher than those of the damped states which indicates that our recovery scheme also works for the two-qubit mixed states. To justify whether our method works for general two-qubit mixed states, we also perform the numerical calculation of average fidelity of the damped states and the recovered states over a large ensemble. To do so, we randomly generate a large ensemble of two-qubit mixed state using the method shown in \cite{miszczak2012generating,zyczkowski2011generating,mezzadri2006generate} in which they would obey the required properties of a valid density matrix from a certain probability distribution. For each decaying probability $p$, $\theta$ is chosen to be $tan^{-1}(1/\sqrt{1-p})$ and the average fidelity of the damped and recovered states are shown in Fig.~\ref{fig-fidelity-Unknown} where we can see that our recovery scheme can effectively restore the general two-qubit mixed state.

\subsection{Entanglement protection from amplitude damping}\label{sec-concurrence}

In this subsection, we study whether the quantum entanglement of the two-qubit mixed state can be protected by our scheme or not. The quantum entanglement of a two-qubit mixed state can be calibrated by the quantum \emph{``concurrence''} which is defined as~\cite{wootters1998entanglement}
\begin{equation}\label{eqn-concurrence}
C(\rho )\equiv \max { (0,{ \lambda  }_{ 1 }-{ \lambda  }_{ 2 }-{ \lambda  }_{ 3 }-{ \lambda  }_{ 4 }) } ,
\end{equation}
in which ${ \lambda  }_{ 1 },...,{ \lambda  }_{ 4 } $ are the eigenvalues in decreasing order of the Hermitian matrix $R(\rho )=\sqrt { \sqrt { \rho  } \tilde { \rho  } \sqrt { \rho  }  }$ 
with $\tilde { \rho  } =({ \sigma  }_{ y }\bigotimes { \sigma  }_{ y }){ \rho  }^{ * }({ \sigma  }_{ y }\bigotimes { \sigma  }_{ y })$.

\begin{figure}
\begin{center}
\subfigure[ Concurrence for density matrix $\rho_1$]{\label{fig-concurrence1}
\includegraphics[scale=0.5]{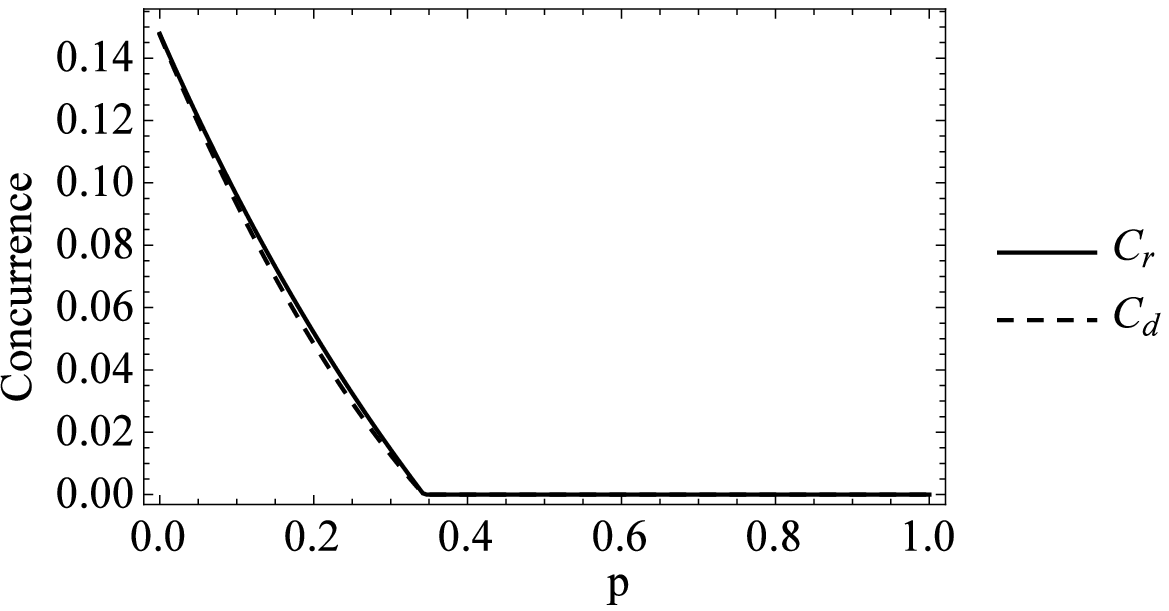}}
\subfigure[Concurrence for density matrix $\rho_2$]{\label{fig-concurrence2}
\includegraphics[scale=0.5]{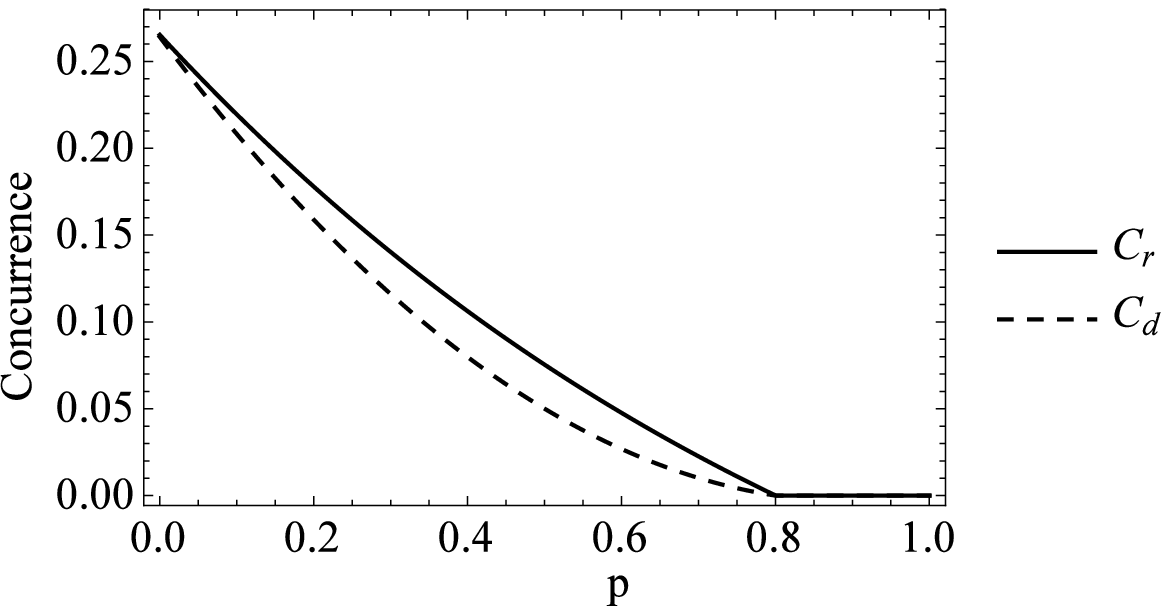}}
\caption{Concurrence as a function of damping probability, $p$, for damped state and recovered state. Corresponding to $\rho_1$ and $\rho_2$ described in Figure \ref{fig-Fidelity1} and \ref{fig-Fidelity2} respectively.}
\end{center}
\end{figure}

The damped concurrence and the recovered concurrences for the two quantum states used in the previous section are shown in Fig.~\ref{fig-concurrence1} and \ref{fig-concurrence2} respectively.
From Fig.~\ref{fig-concurrence1} and \ref{fig-concurrence2} one can see the following features: (1) The concurrence of recovered state is higher than that of the damped one which indicates that our scheme can protect the quantum entanglement of the two-qubit mixed state from amplitude damping. However, the amount of the quantum entanglement does not improve very much. (2) The entanglement vanishes at a special point which is called \emph{entanglement sudden death} (ESD)~\cite{tahira2008entanglement, yu2006sudden}. Before the ESD point, the quantum entanglement can be restored by a certain amount. However, beyond the ESD point, the quantum entanglement can not be improved by the quantum algorithm shown in Fig.~\ref{fig-recovery} because the recovering scheme shown in Fig. 1 is essentially non-unitary local operation.

\section{Extended scheme}\label{sec-extended}

\begin{figure}
\begin{center}
\includegraphics[scale=0.5]{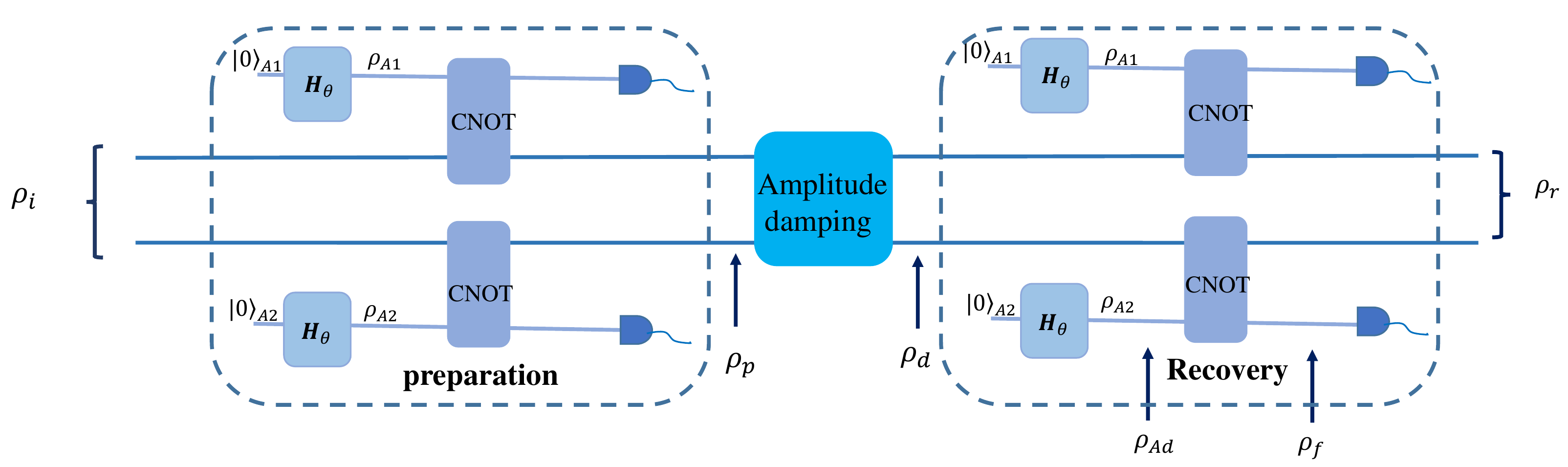}
\caption{A schematic view of the extended scheme process proposed in~\cite{liao2013protecting}, generalized herein for the mixed states setting.}
\label{fig-extend}
\end{center}
\end{figure}

In the previous section, we show that a quantum state can be recovered with very good fidelity by the scheme shown in Fig. 1. However, the quantum entanglement can not be well recovered in that scheme, especially if the ESD occurs by the amplitude damping. In this section, we discuss how to improve this scheme.  Similar to that of~\cite{liao2013protecting}, we can significantly improve the fidelity and quantum entanglement by adding a preparation stage before the amplitude damping of a two-qubit mixed state. The extended scheme scenario is depicted in Fig.~\ref{fig-extend}, which is a mixed-state generalization of the scheme in~\cite{liao2013protecting}.

\begin{figure}
\begin{center}
\subfigure[Density matrix $\rho_1$]{\label{fig-extendeds-cheme1}
\includegraphics[scale=0.5]{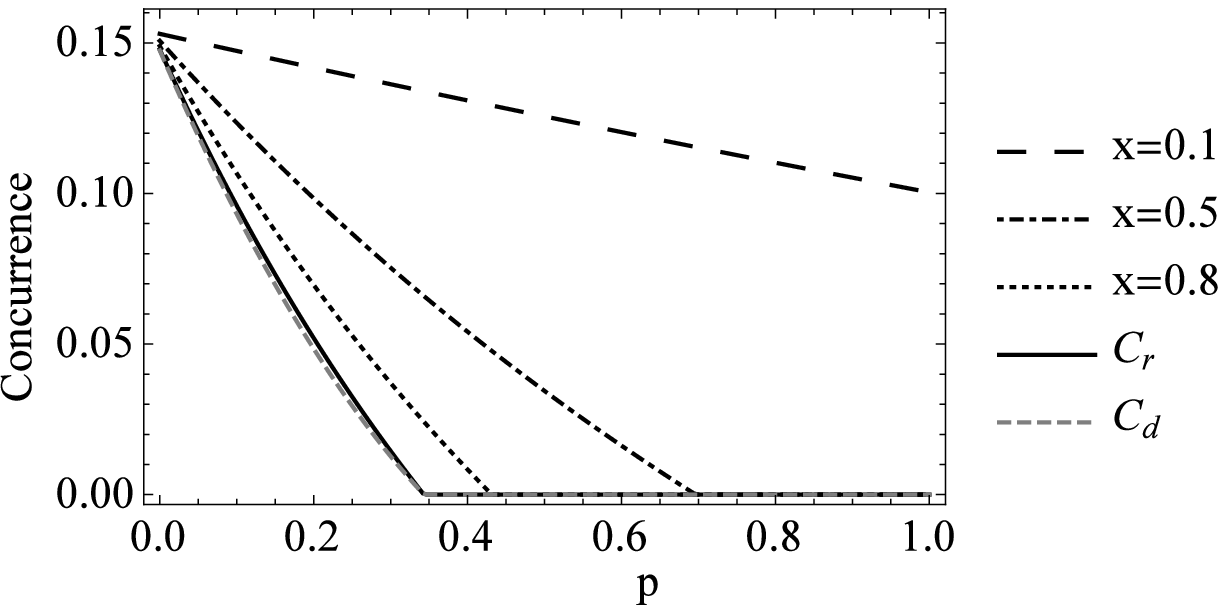}}
\subfigure[Density matrix $\rho_2$]{\label{fig-extendeds-cheme2}
\includegraphics[scale=0.5]{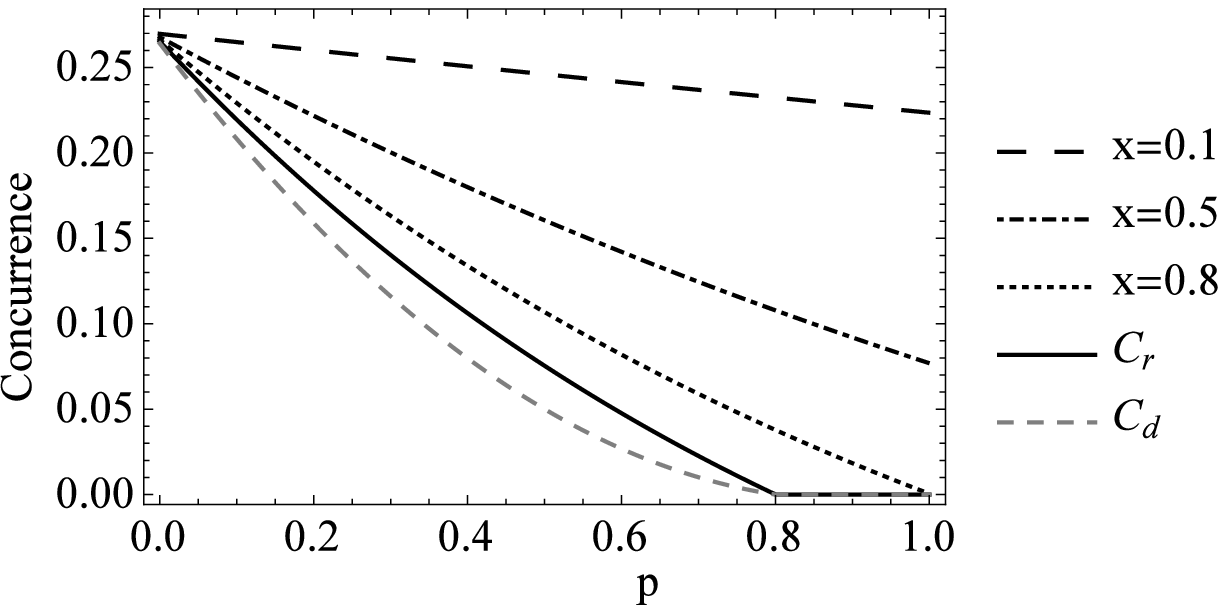}}
\caption{Concurrence as a function of damping probability for damped state and recovered state corresponds to the results in Section~\ref{sec-concurrence} and the other curves relates to $x=0.1$, $x=0.5$ and $ x=0.8$. All curves are belonging to $\rho_1$ and $\rho_2$ described in Figure \ref{fig-Fidelity1} and \ref{fig-Fidelity2} respectively.}
\end{center}
\end{figure}

The proposed method proceeds as follows: Before the initial two-qubit mixed states undergoes amplitude damping, we pre-process the system to make it robust against the amplitude damping. To do so, we apply the same quantum circuit as in the recovery part to \emph{prepare} the initial state. In this stage, the preparation is successful if the ancilla qubits are measured to be $\ket{00}$. After the preparation stage, the system undergoes the damping stage shown in Sec. \ref{sec-damping}. In the final part, we perform the same recovery procedure as shown in Sec.~\ref{sec-recovery} to recover the quantum state and quantum entanglement. 

The quantum state after the preparation stage can be obtained from Eq. \eqref{eqn-recovered}, by considering $p = 0$ and $\theta={\theta}_{1}$ where ${\theta}_{1}$ is the rotation angle of Hadamard gate in the preparation step. Then, by denoting $x\equiv \tan^{2}{\theta}_{1}$, the quantum state after the preparation stage is given by 
\begin{equation}\label{eqn-intialprep}
{ \rho  }_{ p }=\left( \begin{array}{cccc}\frac { a }{ (1+x)^{ 2 } } &{e}{ (\frac { 1 }{ 1+x } ) }^{ \sfrac { 3 }{ 2 }  }\sqrt { \frac { x }{ 1+x }  } & f{ (\frac { 1 }{ 1+x } ) }^{ \sfrac { 3 }{ 2 }  }\sqrt { \frac { x }{ 1+x }  } & \frac { gx }{ { (1+x) }^{ 2 } }  \\  {e}^{*}{ (\frac { 1 }{ 1+x } ) }^{ \sfrac { 3 }{ 2 }  } \sqrt { \frac { x }{ 1+x }  }& \frac { bx }{ { (1+x) }^{ 2 } } & \frac { hx }{ { (1+x) }^{ 2 } } & i{ (\frac { 1 }{ 1+x } ) }^{ \sfrac { 3 }{ 2 }  }\sqrt { \frac { x }{ 1+x }  }\\  {f}^{*}{ (\frac { 1 }{ 1+x } ) }^{ \sfrac { 3 }{ 2 }  } \sqrt { \frac { x }{ 1+x }  } &\frac {{ h}^{*}x }{ { (1+x) }^{ 2 } }&  \frac { cx }{ { (1+x) }^{ 2 } }&j{ (\frac { 1 }{ 1+x } ) }^{ \sfrac { 3 }{ 2 }  }\sqrt { \frac { x }{ 1+x }  }\\ \frac { {g}^{*}x }{ { (1+x) }^{ 2 } }&{i}^{*}{ (\frac { 1 }{ 1+x } ) }^{ \sfrac { 3 }{ 2 }  } \sqrt { \frac { x }{ 1+x }  }&{j}^{*}{ (\frac { 1 }{ 1+x } ) }^{ \sfrac { 3 }{ 2 }  } \sqrt { \frac { x }{ 1+x }  }&\frac { dx^2 }{ { (1+x) }^{ 2 } }  \end{array} \right) .
\end{equation}

It is noted that if ${\theta}_{1}$ is selected such that $x<1$, the system uncollapses toward the ground state as weak measurement~\cite{korotkov2010decoherence,kim2012protecting}. The ground state is less vulnerable to the amplitude damping because it is uncoupled to the environment~\cite{korotkov2010decoherence}. In the next stage, the prepared state shown in Eq. \eqref{eqn-intialprep} undergoes the amplitude damping and the recovery procedure, shown in Fig \ref{fig-extend}. For the recovery stage we determine the rotation angle of the Hadamard gate, ${\theta}_{2}$, such that $xqy=1$ where $ y\equiv\tan^{2}{\theta}_{2}$. Then, as in Sec.~\ref{sec-recovery}, we measure the ancilla qubits in $\ket{00}$ states, and finally obtain the recovered density matrix.

\begin{figure}
\begin{center}
\subfigure[Fidelity for density matrix $\rho_1$]{\label{fig-extendedfidelity1}
\includegraphics[scale=0.5]{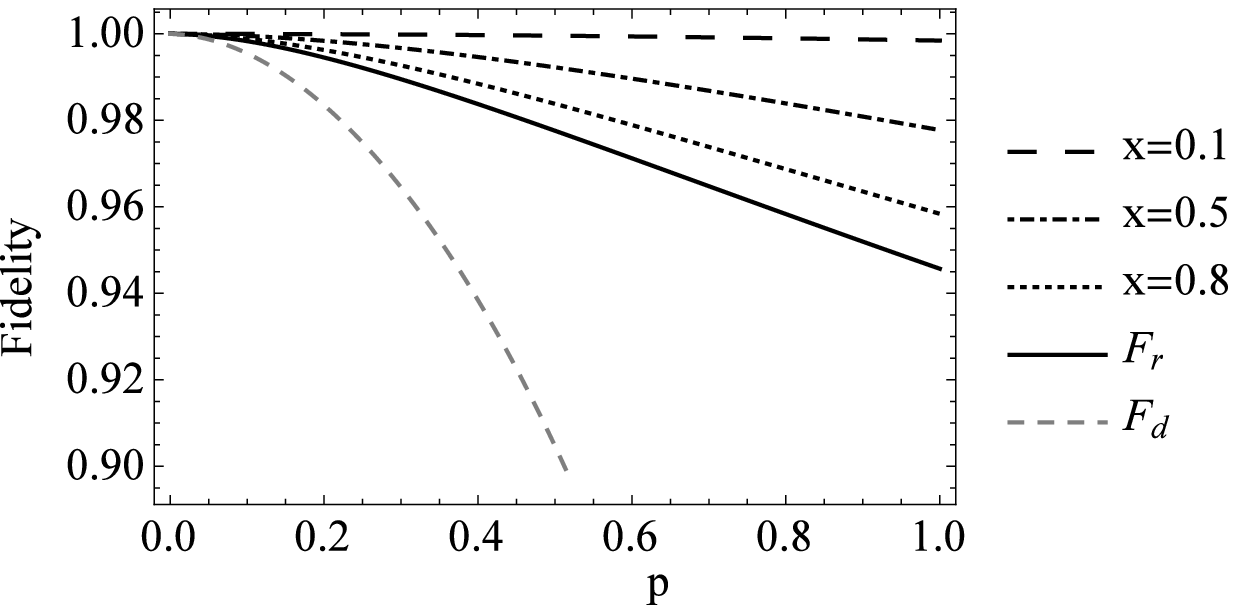}}
\subfigure[Fidelity for density matrix $\rho_2$]{\label{fig-extendedfidelity2}
\includegraphics[scale=0.5]{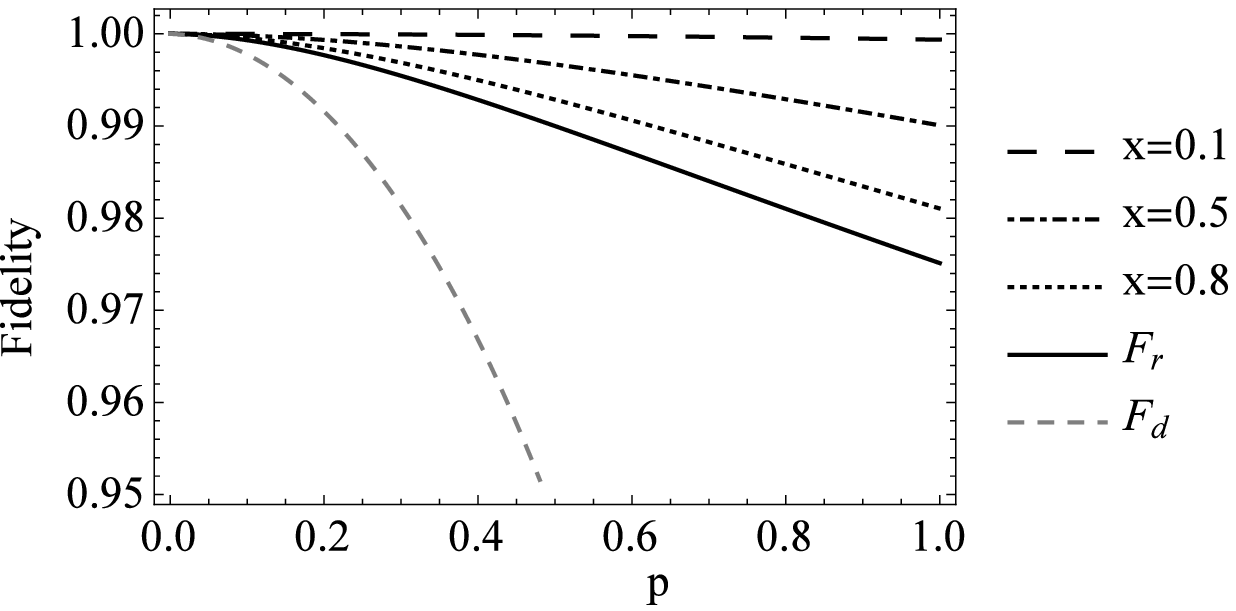}}
\caption{Fidelity in the extended scenario as a function of damping probability. Damped state and recovered state corresponds to the results in Section~\ref{sec-fidelity}. The other curves relates to $x=0.1$, $x=0.5$ and $ x=0.8$ in the extended scenario. All curves are belonging to $\rho_1$ and $\rho_2$ described in Figure \ref{fig-Fidelity1} and \ref{fig-Fidelity2} respectively.}
\end{center}
\end{figure}

We now examine how our extended scheme works compared with the scheme without preparation stage. In Fig. \ref{fig-extendeds-cheme1} and \ref{fig-extendeds-cheme2}, we show the quantum entanglement recovery ${C}_{r}$ under different values of $x$. From the figures, we see that ${C}_{r}$ in the extended scheme with $x<1$ can be higher than $C_{r}$ in the previous scheme without preparation stage. When $x=1$, ${C}_{r}$ in the extended scheme returns back to the previous one. In addition, we notice that the quantum entanglement does not vanish in the extended scheme even beyond the ESD point which never occur in the previous scheme. The fidelity of the recovered state can be also significantly improved in the extended scheme (see Fig. \ref{fig-extendedfidelity1} and  \ref{fig-extendedfidelity2}). However, we should note that the success probability decreases when $x$ is smaller.

\section{Robust recovery under uncertainty}\label{sec-uncertainty}

In previous sections, we have considered the scenario where we have the complete knowledge about the parameter of the apparatus. It means that our model would work in the situation where we know the exact values of the parameters, e.g. known $p$ and consequently designing $\theta$ based on $p$. In this situation, as described above, we can follow the reversal scheme outlined in Fig.~\ref{fig-recovery}~and~\ref{fig-extend}, and use them to reverse the initial mixed states when it undergoes amplitude damping. 
\\Another question that has been studied is: {\emph{What if we aim to design such an apparatus where we face with some issues of uncertainties?}} One of the important issues is uncertainty on $p$. Furthermore, we know that Hadamard gate angle which works properly for one state may not necessarily be the best one for other states. Below, we depict two scenarios. First, we consider a scenario where we want to design the setup whereas there is a mismatch in the actual $p$ and the one with which we design the angle. We discuss the effect of this mismatching in Sec.~\ref{sec-uncertainty1}. Next, in order to overcome the illustrated shortcoming of this mismatch, we propose a {\emph{robust recovery scheme}} (RRS) where we can find an optimal Hadamard gate angle and it can be indeed helpful for battling against the uncertainty on $p$, and also uncertainty around the input state. This approach would be applicable widely, since it requires no initial assumption. 
\subsection{Uncertainty in $p$}\label{sec-uncertainty1}

In the previous sections, we assume that the decay parameter $p$ is known which led to designing $\theta$ such that $\theta=\tan^{-1}(1/\sqrt{q})$. However, in practice, one may not have a complete estimate of $p$,  i.e. either completely unknown or known upto to an interval. Therefore, a legitimate question can be \emph{``How can we determine the Hadamard gate angle such that given our uncertainty about $p$, the achieved fidelity would become sufficient?''}
\\In order to quantify the degrading effect of an unknown $p$, we conduct a numerical simulation study. Suppose that, we have a point estimation for the value that $p$ can take, say $\hat { p }=0.7$. Then, based on this value, we set the Hadamard gate angle using $\theta=\tan^{-1}(1/\sqrt{1-p})=61.3^{o}$. Now, we are interested in evaluating the fidelity of this ``$\theta$-fixed'' recovery scheme across all the possible actual values of $p$. Moreover, we would like to see the difference in fidelity with the case with known $p$ and adaptive selection of $\theta$ as $\theta=\tan^{-1}(1/\sqrt{1-p})$, i.e. for every $p$ design $\theta$ such that $\theta=\tan^{-1}(1/\sqrt{1-p})$. 
\begin{figure}
\begin{center}
\subfigure[Density matrix $\rho_1$]{\label{fig-fixedtheta1}
\includegraphics[scale=0.48]{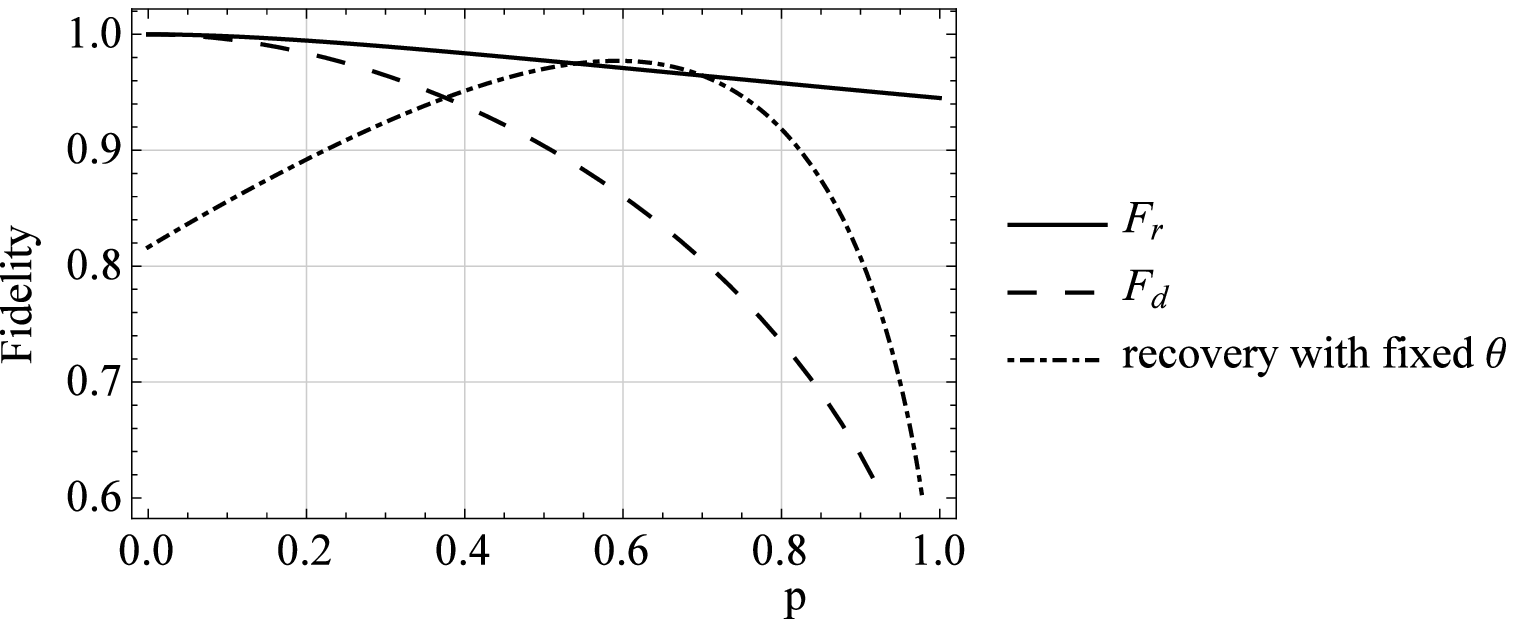}}
\subfigure[Density matrix $\rho_2$]{\label{fig-fixedtheta2}
\includegraphics[scale=0.48]{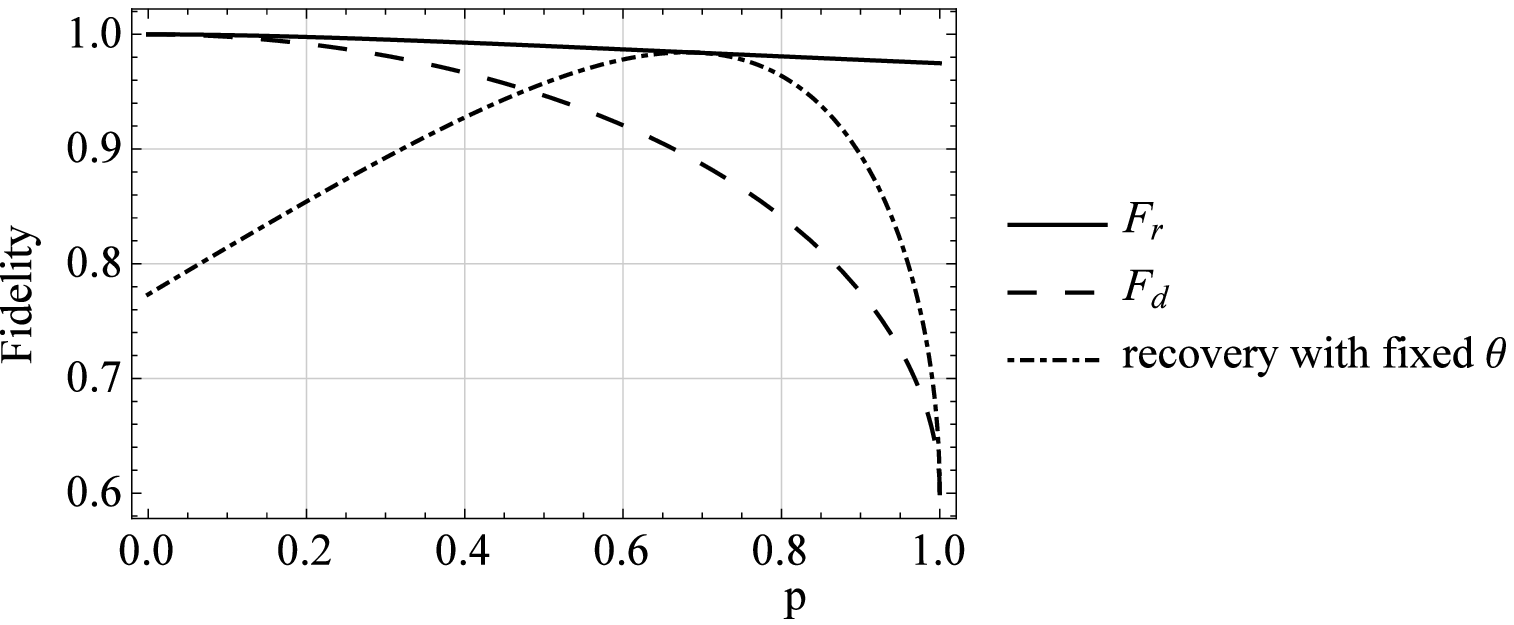}}
\caption{ The system states are $\rho_1$ and $\rho_2$. The solid line and dashed line depends on fidelity with known $p$ and adaptive $\theta$ with $p$, ($\theta=\tan^{-1}(1/\sqrt{1-p}$), used  in \ref{sec-fidelity}. The doted line depends on fixed $\theta$ which obtains from $\theta=\tan^{-1}(1/\sqrt{0.3})$.

}
\end{center}
\end{figure} 
Simulation results, for the two mixed density matrices $\rho_1$ and $\rho_2$ are shown in Figures~\ref{fig-fixedtheta1} and \ref{fig-fixedtheta2}. We plot the fidelity of recovered scheme by considering fixed $\theta$ above (i.e. corresponding to $\hat { p } =0.7$), along with the two other curves, taken from Figs.~\ref{fig-Fidelity1} and \ref{fig-Fidelity2}.

Deducing from Figs. \ref{fig-fixedtheta1} and \ref{fig-fixedtheta2}, we can summarize the simulation results by the following two points:
1) For fixed $\theta=61.3^{o}$, the quantum state is not recovered well on all $p$'s, unless in the range around $p=0.35$ to $1$ for density matrix corresponding to $\rho_1$ and the range around $p=0.55$ to $1$ for $\rho_2$.
2) Even though selection of $\theta$ through the tangent formula shows a better performance overall, as shown in Figs.~\ref{fig-fixedtheta1} and~\ref{fig-fixedtheta2}, one may find a better $\theta$ for a specific damped probability $p$ rather than that calculated by $tan^{-1}(1/\sqrt{1-p})$. 
It seems that in this situation, where we do not know about $p$, choosing a random $p$ to determine the angle $\theta$ is not a considerable way. Hence, in Section~\ref{sec-uncertainty2}, we define a robust method to solve this problem.

\subsection{Unknown $p$ and $\rho$ }\label{sec-uncertainty2}

In the previous subsection, we studied how choosing a Hadamard gate angle $\theta$, where we have uncertainty on $p$, would affect the fidelity under different values of $p$. We want now make our uncertainty broader by assuming uncertainty on both $p$, and the initial quantum state of the system $\rho$. In this scenario, we introduce a robust recovery scheme based on finding an optimal $\theta$ which yields the best average fidelity taken over the distribution. 

\begin{definition}[Fidelity-Robust Recovery Scheme]

Suppose that the (unknown) density states are governed by a given distribution, i.e., each density matrix has also a probability of occurrence. Then, we define fidelity as a function of $\rho$, $p$ and $\theta$, and denote it by $F(\rho ,p,\theta )$. We define the average fidelity over the range of $p$ and $\rho$, as follows
\\\begin{equation}\label{eq-def-expected-fid}
\overline { F } (\theta )={ E }_{ p }\left[ { E }_{ \rho  }\left[ F(\rho ,p,\theta ) \right]  \right] 
\end{equation}
Then, we define a recovery scheme, fidelity-robust, if its Hadamard gate angle ${\theta}_{opt}$ is chosen as follows
\begin{equation}\label{eq-def-max}
{ \theta  }_{ opt }\delequal \max { \overline { F } (\theta ) }.
\end{equation}
We call ${ \theta  }_{ opt }$, the robust Hadamard gate angle. It should be noted that by averaging we cancel out the roles of unknown $\rho$ and $p$ on the fidelity. This is also called marginalization. 
\end{definition}
In the case of a given interval, for the unknown $p\in (p_l,p_u)$, we can simplify Eq. ~\eqref{eq-def-expected-fid}, as follows
\begin{equation}
\overline { F_r } (\theta )=\frac{1}{p_u-p_l}\int_{p_l}^{p_u}  { E }_{ \rho  }\left[ F(\rho ,p,\theta ) \right] dp.
\end{equation}
In many situations, it may not be feasible to find a closed-form solution for either of \eqref{eq-def-expected-fid} or \eqref{eq-def-max}. In these cases, one may take numerical approaches for computing the expectations, and solving the maximization problem. In the following, we show an example, where we find the robust angle via Monte Carlo simulations. 

To examine the performance of the proposed recovery scheme with the robust angle $\theta,$ we compare the case with complete knowledge of $\theta$ (the original scheme) with the $\theta$ obtained from Eq.~\eqref{eq-def-max}:
\begin{equation}
\overline{F}=E_p\Big[E_{\rho}[F(\rho,p,\theta)]|\theta=\tan^{-1}(\sfrac{1}{\sqrt{1-p}})\Big],
 \end{equation}
 
We generate random $\rho$ via Monte Carlo approach with ${10}^4$ iterations, and $p$ also uniformly varies between $0.1$ and $0.9$. To find the maximum average fidelity, we grid the range of $\theta$ between $0$ and $2\pi$ with steps of $\frac { \pi  }{ 10 } $. We summarize the results in Table~\ref{table-results-1}.
 \begin{table}[h]
\centering
\begin{tabular}{l|lllllllllll}
$F$|$\theta$&$ 0$ & $\sfrac { \pi}{ 10 }$ & $\sfrac { 2\pi}{ 10 }$ & $\bf{\sfrac { 3\pi}{ 10 }}$ & $\sfrac { 4\pi}{ 10 }$ & $\sfrac { 5\pi}{ 10 }$ & $\sfrac { 6\pi}{ 10 }$ & $\sfrac {7\pi}{ 10 }$ & $\sfrac { 8\pi}{ 10 }$ & $\sfrac { 9\pi}{ 10 }$ & $\pi$ \\
\hline
$\overline { F } (\theta )$ & $0.250$& $0.427$ & $0.632$ & $\bf{0.790}$ & $0.751$ & $0.248$ & $0.090$ & $0.089$ & $0.114$ & $0.154$ & $0.250$  \\
\hline
$\overline { { F }_{ r } }$ & $\bf{0.838}$ & $0.838$ & $0.838$ & $0.838$ & $0.838$ & $0.838$ & $0.838$ & $0.838$ & $0.838$ & $0.838$ & $0.838$ \\
$\overline { { F }_{ d } } $& $\bf{0.727}$ & $0.727$ & $0.727$ & $0.727$ & $0.727$ & $0.727$ & $0.727$ & $0.727$ & $0.727$ & $0.727$ & $0.727$
\end{tabular}
\caption{Average recovery fidelity for two scenarios. The first row shows the results for the case where $p$ and $\rho$ are assumed unknown. Also the results are repeated periodically until $2\pi$. The second row show the fidelity for the scenario where the gate angle is chosen with the knowledge of $p$ same as Sec. ~\ref{sec-fidelity}. The third row is the average damped fidelity. }
\label{table-results-1}
\end{table}
 \\One can find ${\theta}_{opt}$ (among the selected candidate angles) by choosing the maximum fidelity among these numbers. The optimal fidelity, bolded in the table, is $0.790$ which corresponds to $\sfrac { 3\pi}{ 10 }$. We omit the rest of numbers after $\pi$, because they periodically repeat. Although $0.790$ is smaller than the fidelity $0.838$ in the case we have know exactly what the $p$ is, it is still larger than the damped fidelity $0.727$. One should note that, one may achieve better results by a finer grid of $\theta$.  Therefore, one may conclude that via the robust recovery scheme, the mixed states recovery can be robustly implemented with no knowledge of the underlying $p$ or $\rho$, and we only need to know the distributions governing these two parameters.

\section{Summary}

In summary, we show several schemes to protect an arbitrary two-qubit mixed state from amplitude damping. The basic scheme without preparation stage can recover a quantum state with very high fidelity, but the quantum entanglement of the two-qubit mixed state is not significantly improved. The extended scheme with preparation stage can recover the two-qubit mixed state with a very high fidelity and the quantum entanglement can be also significantly recovered by choosing suitable parameters. Furthermore, the extended scheme can also recover a quantum state even beyond the ESD point.  

In addition, a recovery scheme was next introduced which takes the system's uncertainties into account which in turn led to a robust recovery scheme. We find an optimal angle for recovering a two-qubit mixed state when the quantum state and the decay probability is unknown. This scheme may be very useful for protecting a quantum state from amplitude damping in practice.

\section{Acknowledgment}
This work is supported by a grant from the Qatar National Research Fund (QNRF) under the NPRP project 7-210-1-032.

\section{References}
\providecommand{\newblock}{}

\end{document}